%% file: psrb1259_2021.tex
\documentclass[preprints,article,accept,moreauthors,pdftex]{Definitions/mdpi}

\usepackage{xspace}
\usepackage{comment}
\usepackage{todonotes}
\def\swift{\textit{Swift}\xspace}

\def\xrt{\textit{Swift}/XRT\xspace}

\def\flat{\textit{Fermi/LAT}\xspace}

\def\psrb{PSR~B1259-63\xspace}

\firstpage{1} 
\pubvolume{1}
\issuenum{1}
\articlenumber{0}
\pubyear{2021}
\copyrightyear{2020}
\datereceived{} 
\dateaccepted{} 
\datepublished{} 
\hreflink{https://doi.org/} 

\Title{Multi-wavelength properties of the 2021 periastron passage of \psrb}

\TitleCitation{Title}

\Author{Maria Chernyakova $^{1,2,*}$\orcidA{}, Denys Malyshev $^{3}$\orcidE{}, Brian van Soelen $^{4}$, Shane O'Sullivan $^{1}$\orcidD{}, Charlotte Sobey $^{5}$\orcidS{}, S. Tsygankov $^{6,7}$\orcidT{}, Samuel Mc. Keague $^{1}$, Jacob Green $^{1}$, Matthew Kirwan\orcidB{} $^{1}$ , Andrea Santangelo$^3$, Gerd P\"uhlhofer$^3$, Itumeleng M. Monageng$^{8,9}$}

\AuthorNames{Firstname Lastname, Firstname Lastname and Firstname Lastname}

\AuthorCitation{Chernyakova, M.; Malyshev, D.; van Soelen, B. et al}

\address{%
$^{1}$ \quad School of Physical Sciences and Centre for Astrophysics \& Relativity, Dublin City University, Glasnevin, D09 W6Y4, Ireland.\\
$^{2}$ \quad Dublin Institute for Advanced Studies, 31 Fitzwilliam Place, Dublin 2; \\
$^{3}$ \quad Institut f{\"u}r Astronomie und Astrophysik T{\"u}bingen, Universit{\"a}t T{\"u}bingen, Sand 1, D-72076 T{\"u}bingen, Germany\\ 
$^{4}$ \quad Department of Physics, University of the Free State, PO Box 339, Bloemfontein 9300, South Africa \\
$^{5}$ \quad CSIRO Astronomy and Space Science, PO Box 1130, Bentley, WA 6102, Australia\\
$^{6}$ \quad Department of Physics and Astronomy, FI-20014 University of Turku,  Finland\\
$^{7}$ \quad Space Research Institute of the Russian Academy of Sciences, Profsoyuznaya Str. 84/32, Moscow 117997, Russia\\
$^{8}$ \quad South African Astronomical Observatory, PO Box 9, Observatory, 7935, Cape Town, South Africa\\ 
$^{9}$ \quad Department of Astronomy, University of Cape Town, Private Bag X3, Rondebosch 7701, South Africa\\
}

\corres{e-mail: masha.chernyakova@dcu.ie }

\abstract{\psrb is a gamma-ray binary system hosting a radio pulsar orbiting around a O9.5Ve star, LS 2883, with a period of $\sim$ 3.4 years. The interaction of the pulsar wind with the LS 2883 outflow leads to unpulsed broadband emission in the radio, X-ray, GeV, and TeV domains. One of the most unusual features of the system is an outburst at GeV energies around the periastron, during which the energy release substantially exceeds the spin down luminosity under the assumption of the isotropic energy release. 
In this paper, we present the first results of a recent multi-wavelength campaign (radio, optical, and X-ray bands) accompanied by the analysis of publicly available GeV \flat data. The campaign covered a period of more than 100~days around the 2021 periastron and revealed substantial differences from previously observed passages. We report a major delay of the GeV flare, weaker X-ray flux during the peaks, which are typically attributed to the times when the pulsar crosses the disk, and the appearance of a third X-ray peak never observed before. We argue that these features are consistent with the emission cone model of Chernyakova et al (2020) in the case of a sparser and clumpier disk of the Be star.}


\begin{document}


\section{Introduction}

\psrb (PSR J1302--6350) is a 47.76-ms radio pulsar orbiting an O9.5Ve star (LS~2883) with a period of $\sim 1236.7$~days in a highly eccentric orbit ($e\sim 0.87$) \citep{1992ApJ...387L..37J,Negueruela-2001,shannon14}. The Be stellar disk is highly inclined to the orbital plane~\citep{Wang2004}, so that the pulsar crosses the disk twice per orbit. Based on the parallax data in the Gaia DR2 Archive~\citep{Gaia2018} the distance to the system is $2.39 \pm 0.19$~kpc, which is consistent with the value of $2.6^{+0.4}_{-0.3}$ kpc reported previously  \cite{PSRB1259-2018_distance}. \psrb is one of two known systems where an interaction between the pulsar wind and the mass outflow from the young, massive companion leads to broadband (from radio up to TeV energies), unpulsed non-thermal emission (see e.g. \cite{Chernyakova2020review}, and references therein). Another known pulsar with similar broadband properties, PSR J2032+4127, has an orbital period of about 50 years \citep{Ho2017,Chernyakova2020psrj} and thus the possibilities to study the details of the interaction of the winds near the periastron in this system are very limited.

\psrb was discovered in 1992 during the Parkes Galactic plane survey \citep{1992ApJ...387L..37J,1992MNRAS.255..401J} as the first radio pulsar in orbit around an OB star. Further observations revealed that the pulsed radio emission is completely absorbed from around 20 days before to 20 days after periastron, but that unpulsed radio emission appears about 20 days before the periastron, quickly rising to a flux exceeding the pulsed flux by a factor of several tens, and shows two peaks that are associated with the pulsar crossing the plane of the circumstellar disk, before and after the periastron, and lasts for at least 100 hundred days after periastron \citep{2002MNRAS.336.1201C,2005MNRAS.358.1069J}. In X-rays, \psrb is visible throughout the whole orbit with two flux peaks approximately 15 days before and after periastron, similar to the radio light curve \citep[see e.g.][]{Chernyakova2006,Chernyakova2009}.

During the periastron passage, \psrb is also visible at high and very high energies \citep[e.g.][]{FERMI_PSRB2018,hess_psrb2020}. In 2010, 2014 and 2017, the light curve at GeV energies was marked by the presence of a strong flare, which occurred about 30 - 40 days after the periastron \citep[e.g.][]{2011ApJ...736L..11A,caliandro15,FERMI_PSRB2018}. In 2017, the flare was characterized by the presence of variability on time-scales as short as 
15 minutes, during which the energy release substantially exceeded the spin down luminosity (in the case of an isotropic outburst) by a factor of $\sim30$ \citep{Tam2018,FERMI_PSRB2018}. These flares were observed only at GeV energies with no obvious counterparts at other wavelengths despite the possible link to a decrease of the H$\alpha$ equivalent width \citep{2014MNRAS.439..432C,chernyakova15}. To explain the luminosity of the GeV flare and the absence of counterparts at other wavelengths, a model was proposed \cite{Chernyakova20_psrb} in which the TeV and X-ray emission is generated by the strongly accelerated electrons of the pulsar wind (IC and synchrotron emission correspondingly). The GeV emission in this model is a result of the IC emission of the unshocked and weakly shocked electrons, with a possible addition of bremsstrahlung emission on the clumps of the stellar wind material which penetrated beyond the shock cone. The unshocked electrons of the pulsar wind in this model are quasi-monochromatic, with energies of about 1 GeV. The luminosity of the GeV flares within this model can be understood if it is assumed that the initially isotropic pulsar wind after the shock is reversed and confined within a cone looking, during the flare, in the direction of the observer.

In order to test this model, we organised an intensive multi-wavelength campaign to follow the 2021 periastron passage of \psrb in radio (ATCA), optical (SALT) and X-rays (\xrt). Below we present the first results of this campaign, including the analysis of the GeV data as seen by the Large Area Telescope (LAT) on board the Fermi Gamma-ray Space Mission. In section 2 of this paper, we describe the details of the observations and data analysis, in section 3, we present our results, and we give our conclusions in section 4.

\section{Data Analysis}

\begin{figure}
\includegraphics[width=\columnwidth]{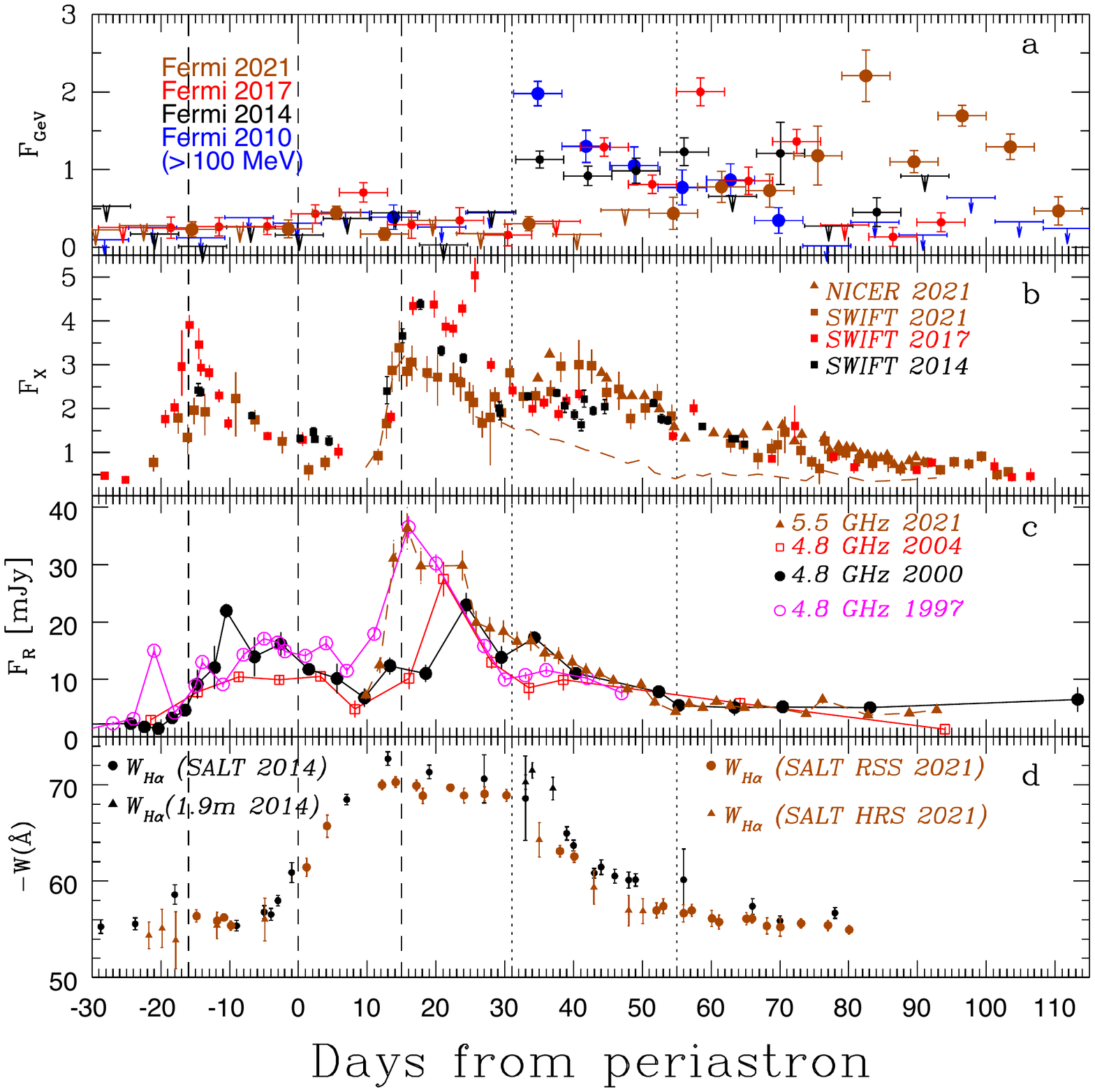}
\vspace{-3.5cm}
\caption{Evolution of \psrb flux over the different periastron passages.Dashed lines correspond to the periastron and to the moments of
disappearance (first non-detection) and reappearance (first detection) of the pulsed emission, as observed in 2010 \citep{2011ApJ...736L..11A}. Dotted lines correspond
to the first appearance of the detection in GeV band at a day time scale in 2010 and 2021. \textit{Panel a:} \flat flux measurements in the E $>$ 100 MeV energy range with a weekly bin size. Flux is given in 10$^{-6}$ cm $^{-2}$ s $^{-1}$. \textit{Panel b:} absorbed 1-10 keV X-ray flux in units of 10$^{-11}$ erg cm $^{-2}$ s $^{-1}$. Scaled 5.5-GHz radio data from 2021 are also shown in this panel with a gold dashed line for comparison. \textit{Panel c:} radio flux densities in mJy. \textit{Panel d:} H\,$\alpha$ equivalent width. 
 }
\label{fig:lc}
\end{figure}

\subsection{Radio Data}

The Australia Telescope Compact Array (ATCA) was used to obtain the radio light curve of \psrb from 19 Feb 2021 to 13 May 2021. The ATCA is a radio interferometer with six 22-m antennas providing a maximum baseline of 6 km, located in Narrabri, NSW, Australia. The observations of \psrb were conducted in the 4-cm band using the Compact Array Broadband Backend (CABB) with centre frequencies of 5.5 GHz and 9 GHz, each with 2 GHz of bandwidth and 1 MHz frequency channels \citep{wilson2011}.
CABB was configured in a mode that allowed us to record the radio continuum and pulsar-binning data.

We observed \psrb for 3 hours approximately every 2 days (as allowed by the telescope schedule) across the above time period. The ATCA primary calibrator B1934-638 was observed for at least 10 mins during each observation. We observed a secondary calibrator, J1322-6532, for 2 mins for every 20 mins on the target (i.e.~\psrb). Here we report on the continuum data analysis at 5.5 GHz only. Later work will include a full analysis of the data at both centre frequencies, providing information on the spectral, polarization and Faraday rotation measure behavior of both the continuum emission and the pulsed emission. 

The Miriad software was used for the data calibration and analysis, following standard routines \citep{sault1995}. The bandpass calibration and the absolute flux density calibration were done using B1934-638. The bandpass calibration was copied to the secondary (phase) calibrator, J1322-6532, which has an angular separation of 2.7 deg from the target. The J1322-6532 data were then used to calibrate the frequency-dependent complex gains and their time variations, as well as the instrumental polarization. Before copying these solutions to the target, the flux density of J1322-6532 was scaled to that of B1934-638. Flagging of radio frequency interference (RFI) and time-ranges with poor data quality was done in an iterative manner during the above process. The calibration routines described above were then repeated to obtain the final calibrated visibility data. Since the target is a point source at the phase centre, the 5.5 GHz flux density values were obtained using the mean visibility value reported by the Miriad task {\sc uvflux}, with the uncertainty in this value obtained by dividing the quoted rms scatter in the visibilties by the square root of the number of correlations. 

\subsection{Optical Data}

Spectroscopic observations were undertaken around periastron using the Southern African Large Telescope (SALT), with both the RSS and HRS spectrographs \cite{buckley06,burgh03,bramall10}. Between 22 days before, to 80 days after periastron, the source was observed 28 times with the RSS using the pg0900 and pg2300 grating, and 10 times with the HRS in Medium Resolution mode ($R = 40000$). The RSS data was pre-reduced using the SALT pipeline, and flat correction, wavelength calibration, and spectral extraction was performed using {\sc numpy/scipy/astropy}. 
The HRS observations were reduced using the pipeline presented in \cite{kniazev06}. Individual spectra taken on the same night were averaged, continuum corrected, and barycentric corrected. We report here on the change in the equivalent width of the H\,$\alpha$ line, and future publications will present a more detailed analysis of the optical behavior. The equivalent width for each observation was measured multiple times, randomly shifting the wavelength range by $\pm4$\,\AA{}, and the median and standard deviation of the measurements obtained are reported as the value and uncertainty. The choice of continuum correction of the observations can introduce additional systematic offsets and we ensured that the continuum correction used, results in the equivalent width measured with the RSS and HRS on 2021-01-28 ($\approx12$\,days before periastron) having, within error, the value. The H\,$\alpha$ equivalent width is shown in the lower panel of Fig.~\ref{fig:lc}.

\subsection{X-ray Data}
\subsubsection{\xrt} 
A full overview of the X-ray flux for the SWIFT observations around 2014, 2017 and 2021 years are presented in the panel (b) of Fig. \ref{fig:lc}. Historical data in this figure are taken from \cite{Chernyakova20_psrb}.
The 2021 periastron passage of \psrb was closely monitored by the \swift satellite \citep{2004ApJ...611.1005G}. We have analyzed all available data taken from January, 19th, 2021 to May, 24th, 2021.  The data was reprocessed and analysed as suggested by the \xrt team \footnote{see e.g. the \href{https://swift.gsfc.nasa.gov/analysis/xrt_swguide_v1_2.pdf}{\xrt User's Guide}} with the \texttt{xrtpipeline v.0.13.5} and \texttt{heasoft v.6.28} software package. The spectral analysis of \xrt spectra was performed with \texttt{XSPEC v.12.11.1}. The spectrum was extracted from a circle of radius $36''$ around the position of \psrb, and the background estimated from a co-centred annulus with inner/outer radii of $60''/300''$. We performed the fit of the spectrum, grouped to have at least 1 photon per energy bin using cash statistic~\citep{cash79} by an absorbed powerlaw model (\texttt{cflux*tbabs*po}) in 0.3-10~keV range. The flux of the source, hydrogen column density and the slope of the powerlaw were treated as free parameters during the fit. The uncertainties of \xrt flux shown in Fig.~\ref{fig:lc} are $1\sigma$ c.r.

\subsubsection{NICER}

About 35 days after the periastron passage, the {\it NICER} instrument \citep{2012SPIE.8443E..13G} started to monitor \psrb's evolution. The {\it NICER} data was reduced using the NICERDAS software version 2020-04-23\_V007a with default filtering criteria applied. The background was estimated using the {\it NICER} tool {\sc nibackgen3C50}\footnote{\url{https://heasarc.gsfc.nasa.gov/docs/nicer/tools/nicer_bkg_est_tools.html}} (Remillard et al., in prep.) with the default parameters. The spectra obtained for each observation were binned to have at least 1 count per energy bin, and W-statistic\footnote{\url{https://heasarc.gsfc.nasa.gov/xanadu/xspec/manual/XSappendixStatistics.html}} \citep{1979ApJ...230..274W} was applied while fitting the data in the 0.5-10 keV energy range. The relatively high background level (around 20-30\% of the total count rate detected from \psrb) did not allowed us to determine reliably the spectral photon index, which is strongly affected by the quality of the background subtraction. Therefore, in order to estimate flux from the source we fixed photon index at averaged value 1.5 obtained from the \xrt data.

\begin{figure}
\includegraphics[width=0.49\columnwidth]{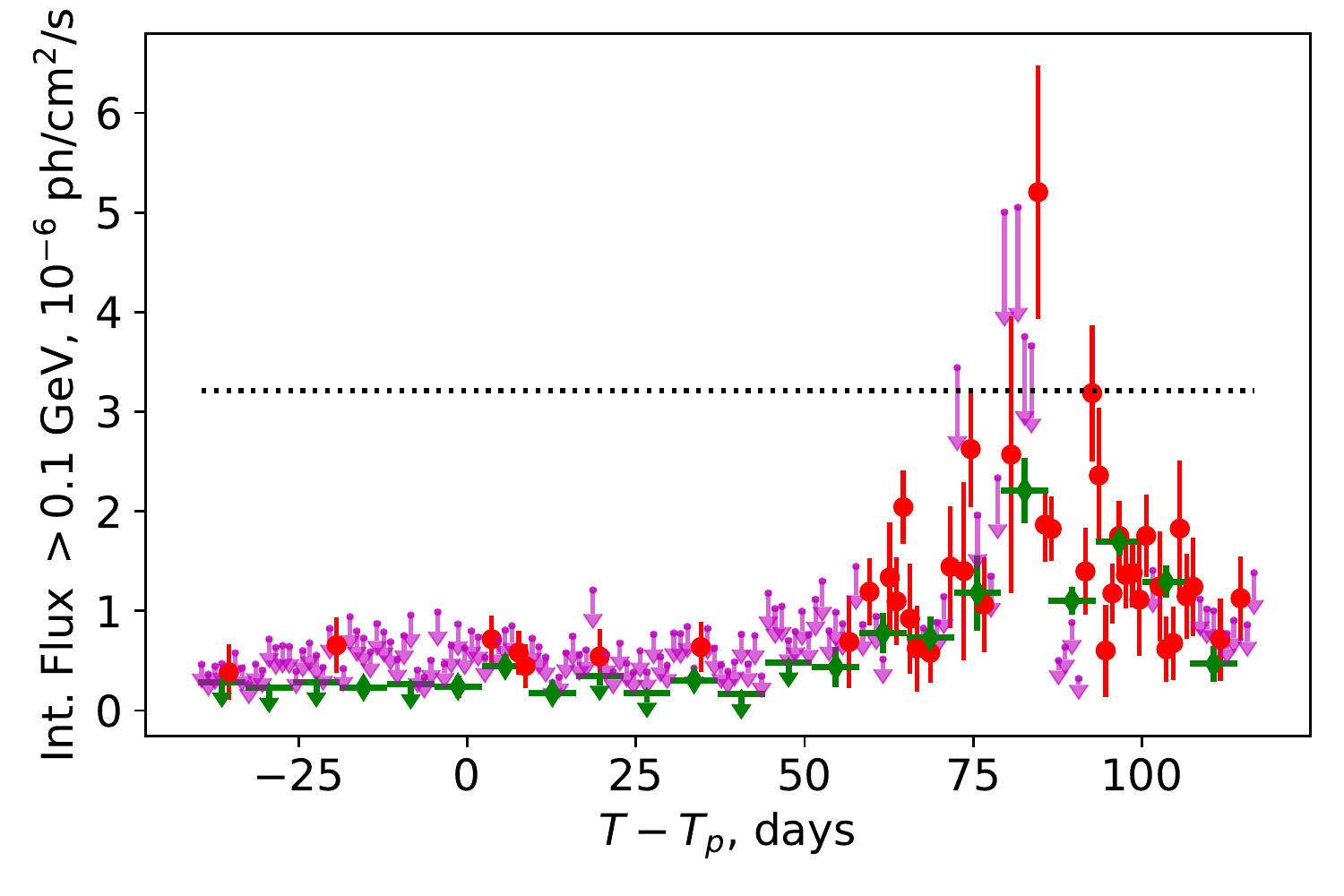}
\includegraphics[width=0.49\columnwidth]{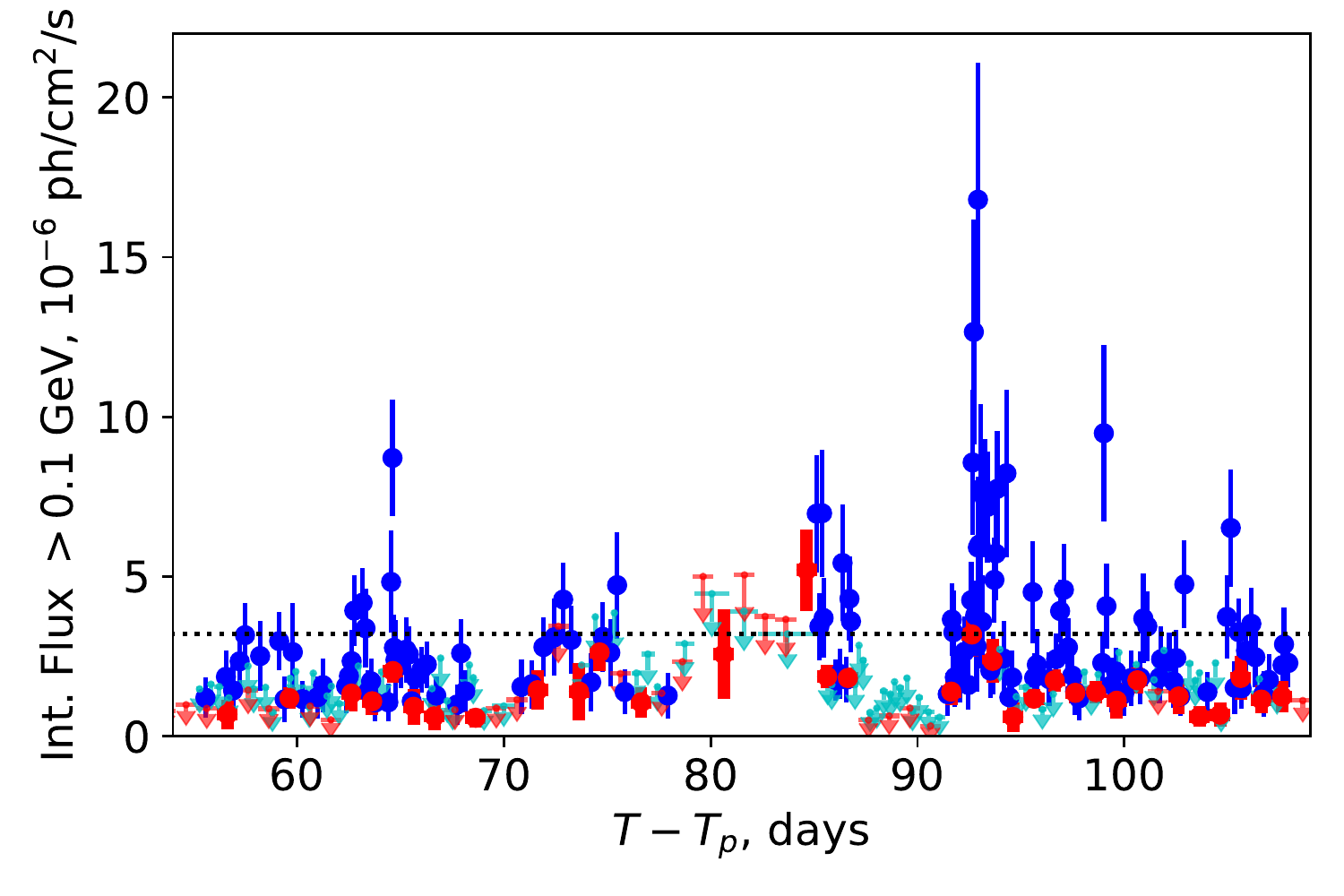}

\caption{Left: \flat weekly (green points) and daily(magenta upper limits and red points -- detection's with $TS>4$) light curve of \psrb at energies $0.1-10$~GeV. The dotted line present the flux corresponding to spin-down luminosity of \psrb ($L_{sd}=8.2\cdot 10^{35}$~erg/s). Right: Blue points: \flat light curve at $>0.1$~GeV range for the period MJD~59310--59363 ((+55; +108)~days after the periastron) with adaptive time binning (to have 9~photons per time bin in $1^\circ$ radius circle around \psrb position at $0.1-10$~GeV). The light curve covers the period of the highest flux of the source in the GeV band. Red points correspond to a daily lightcurve, same as in the left panel. See text for the details.}
\label{fig:fermi_lc}
\end{figure}

\subsection{Gamma-ray data}
The analysis of \flat data was performed using Fermitools version 2.0.8 (released 20th January 2021). For the analysis of the 2021 periastron passage and the combined periastron data, the analysis was carried out using the latest Pass 8 reprocessed data (P8R3) from the CLEAN event class. 
The binned likelihood analysis was performed for photons within the energy range 0.1--100~GeV that arrived between January 1st 2021 and June~6th 2021 within a $20^\circ$-radius region around \psrb's position. The selected maximum zenith angle was $90^\circ$. The performed analysis relies on the fitting of the spectral and spatial model of the region to the observed data in a series of energy and time bins. 

The adapted model of the region included templates for the Galactic and isotropic diffuse emission components provided by \flat collaboration, as well as all sources from the latest 4FGL-DR2 \flat catalogue~\citep{4FGL}, with the spectral templates selected according to the catalogue. At the initial stage of the analysis we assumed all spectral parameters of sources within $15^\circ$ around \psrb to be free parameters and fixed all spectral parameters of sources between $15^\circ$ and $20^\circ$ to their catalogue values. We performed the fitting of the described model to the whole available data-set. For the subsequent analysis, we fixed all (except normalisations) free spectral parameters to their best-fit values, and removed all of the weak sources detected with test-statistics $TS<1$ from the model.

At the next step of the analysis, we split the 0.1-10~GeV data over a series of 1-day and 1-week bins aiming to produce the light curve of \psrb in the corresponding energy band, see Fig.~\ref{fig:fermi_lc}, left panel. All upper limits presented were calculated using the IntegralUpperLimits module included in fermitools for $TS<4$ cases and correspond to a 95\% confidence level.

\begin{figure}
\includegraphics[width=0.49\columnwidth]{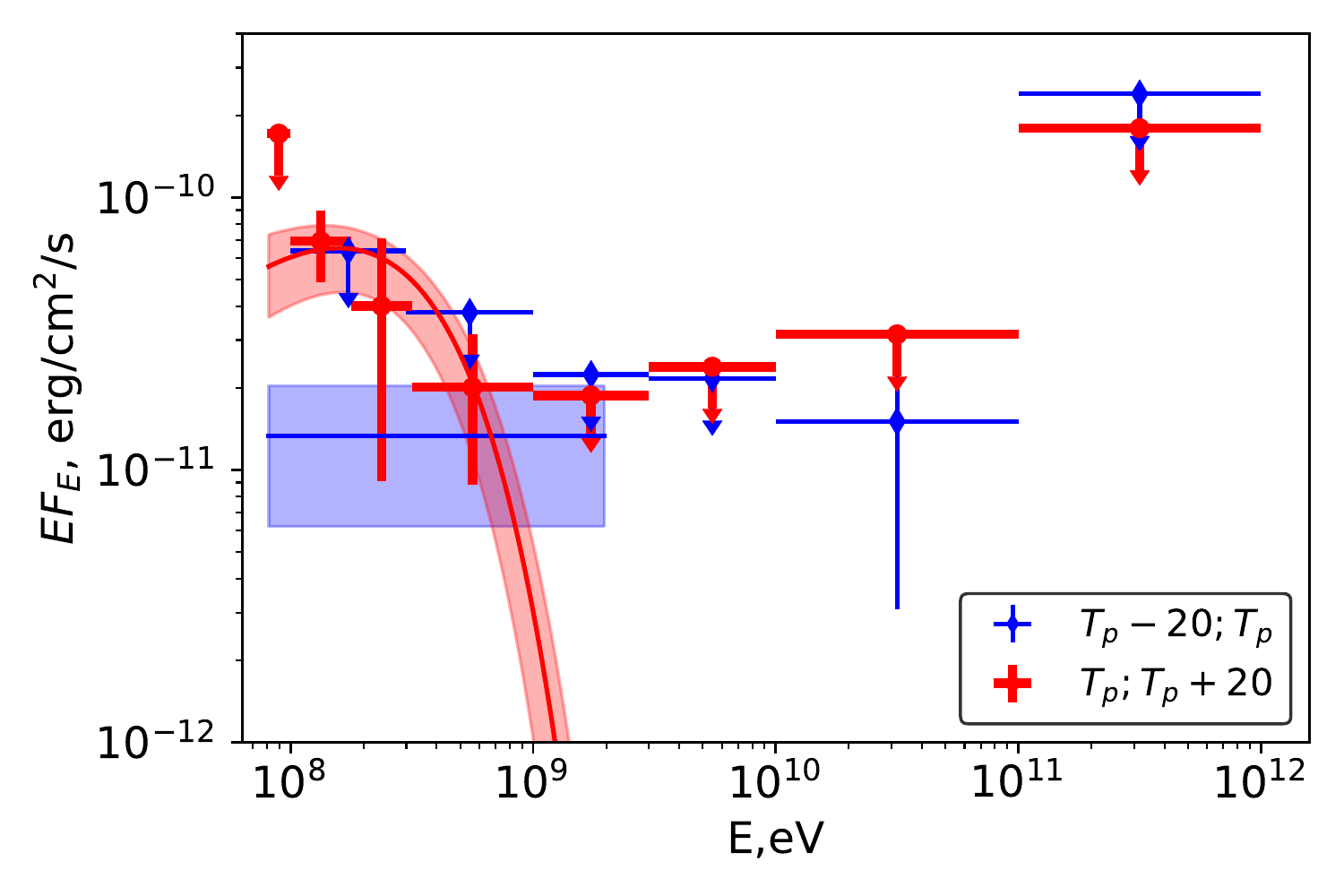}
\includegraphics[width=0.49\columnwidth]{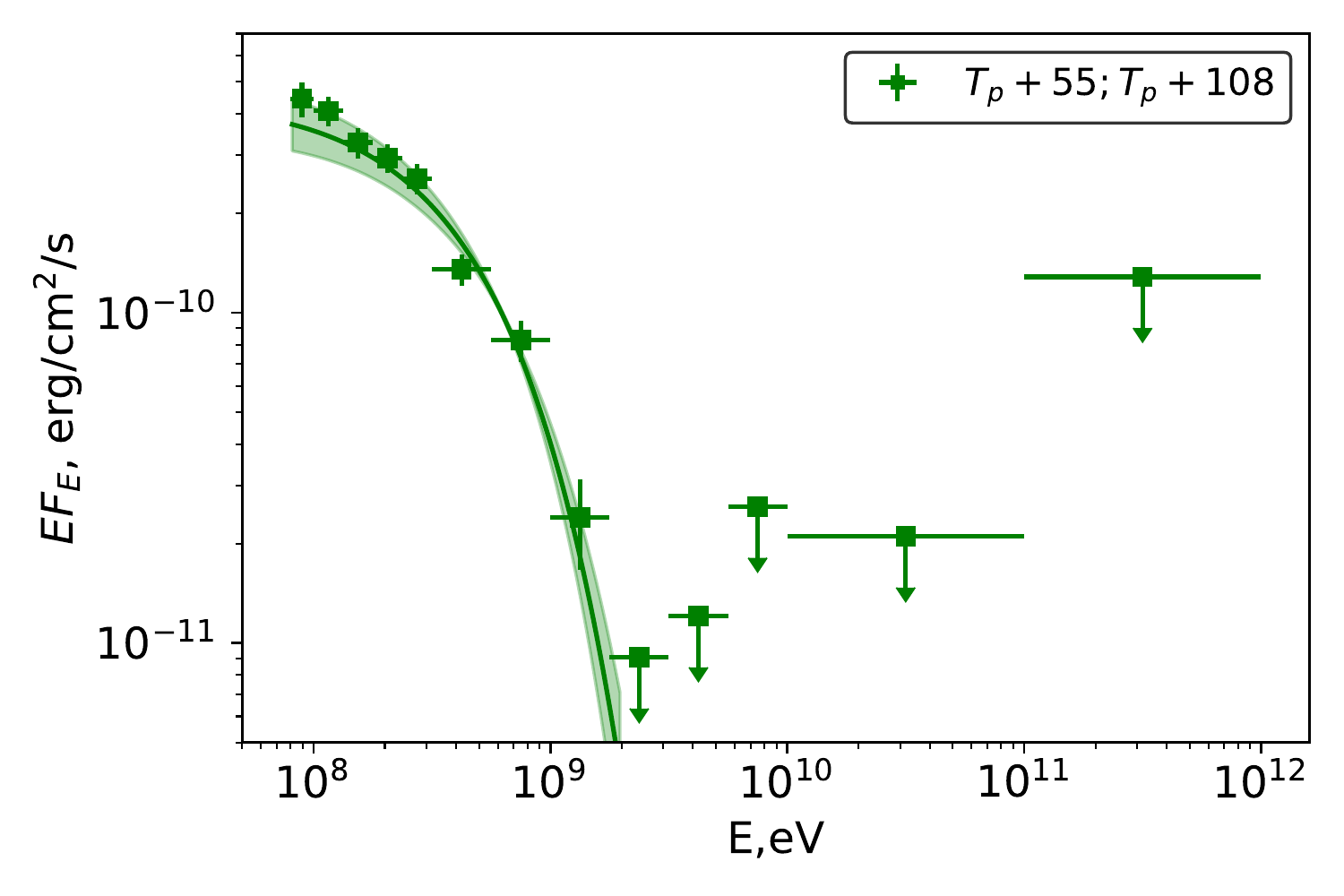}

\caption{Left: \flat spectra of \psrb as seen at (-20 -- 0) days before the periastron (blue points) and (0 -- + 20) days after the periastron (red points). Shaded regions of the corresponding color illustrate $1\sigma$ confidence range for the fitted models. Right: \flat spectrum of the period characterized by the highest flux in $>0.1$~GeV band during 2021 periastron passage (+55 -- +108 days after the periastron), see Tab.~\ref{tab:best_fit_fermi} for the best-fit parameters' values.}
\label{fig:fermi_pm20_spec}
\end{figure}

The obtained lightcurve suggests the enhancement of the GeV emission during $55-108$~days (MJD 59310--59363) after the periastron. In what below we will refer to this period as ``\flat flare'' during 2021 periastron passage. To study the potential spectral-shape variability of the source during different periods of the 2021 periastron passage we built the spectrum of \psrb for the periods $(-20;0)$, $(0; +20)$ days around the periastron, as well as during the \flat flare period. Resulted spectra are shown  in Fig.~\ref{fig:fermi_pm20_spec}. The best-fit spectral parameters in \flat band as well as the suggested for the fit model are summarized in Table~\ref{tab:best_fit_fermi}.

For the period of 2021 \flat flare we performed dedicated studies for a short timescale ($\ll 1$~day) variability aiming to identify bright sub-flares known to be present during 2017 periastron passage. We split the whole time range of \flat flare over a set of variable-length time bins, such that each time bin accommodates 9 photons with energies $0.1-10$~GeV in $1^\circ$-radius circle around \psrb position. The resulted time bins have durations from $5$~min to $2.8$~days with an average duration of $\sim 6$~h. For each of the time bins we extracted the flux of \psrb in a way similar to one which was used for daily/weekly GeV lightcurves production. To produce short timescale lightcurve we explicitly set the spectral model of \psrb to be a (super-exponential)cutoff powerlaw with the indexes and cutoff energy fixed to their best-fit values observed during \flat 2021 flare, see Table~\ref{tab:best_fit_fermi}. The obtained lightcurve is shown in Fig.~\ref{fig:fermi_lc}, right panel.

\begin{table}[]
 \caption{The best-fit parameters of the models (PL -- power law or SECPL -- super-exponential cutoff power law with exponent index $\beta$) used to fit \flat data during different periods of the 2021 periastron passage in $0.08-2$~GeV energy band. See text and Fig.~\ref{fig:fermi_pm20_spec} for the details.}
 \centering
 \begin{tabular}{|ccccccc|}
 \hline
 Period & Model & Norm at 1~GeV & Index & Cutoff& $\beta$ & TS\\
 days & & $10^{-9}$ ph/cm$^2$/s && GeV & & \\\hline
 $(-20; 0)$ & PL & $0.6\pm 0.3$ & 2\,(\textit{fixed})& -- & -- & 7 \\
 $(0; +20)$ & SECPL & $0.5\pm 0.2$ & $1.1\pm 0.2$ & $0.18\pm 0.03$&1\textit{(fixed)} &48 \\
 $(+55; +108)$ &SECPL &$0.31\pm 0.002$& $1.97\pm 0.01$ & $0.38\pm0.01 $& $0.96\pm 0.01$ &1373 \\
 $(-40; +116)$ & PL &$(1.47\pm 0.01)\cdot 10^{-2}$& $2.70\pm 0.01$ & -- & -- &600 \\\hline
\end{tabular}
\label{tab:best_fit_fermi}
\end{table}

\section{Results and Discussion}
The 2021 periastron passage of \psrb is marked by very unusual behaviour both in X-rays and the GeV band. While the rise of the X-ray emission during the first and second disk crossing is similar to that observed during previous periastron passages, 
the heights of these two peaks are substantially smaller (see panel (b) of Figure \ref{fig:lc}). We note that the observations of \psrb in 2021 allow to estimate the characteristic rise time of the X-ray emission during these episodes to be as short as $\sim 1$~day,
 which indicates that the Be star's disk is characterized by high-density gradients.
Most interestingly, the 2021 X-ray light curve shows a third peak, which starts to rise $\approx30$\,days after periastron. This rise occurs at approximately the same time as the GeV flare 
in 2010 and 2014. 
Such X-ray behavior was never observed during previous periastra passages and has no clear counterparts at any other wavelengths.

The 2021 radio observations demonstrate a strong correlation with the second X-ray peak during the period between 10 and 28 days after the periastron. To better illustrate this point, we plot the re-scaled radio data on panel (b) of Figure \ref{fig:lc} (dashed line). With the rise of the third X-ray peak this correlation disappears. At the same time, the 2021 radio observations seem to be consistent with previous observations at 4.8~GHz \citep{2002MNRAS.336.1201C,2005MNRAS.358.1069J}. 

The GeV emission in 2021 is marked by a large delay in the rise of the flux. At early phases (-20 to 0 and 0 to +20 days from periastron) the spectral characteristics of \psrb in the GeV band are consistent with the average values in 2011-2017 reported in \cite{Chernyakova20_psrb}, see Fig.~\ref{fig:fermi_pm20_spec}. However, contrary to previous passages, up to 55~days after periastron, the flux ($>0.1$\,GeV) remained at a level of $(0.2-0.4)\cdot 10^{-6}$ ph/cm$^2$/s, which is consistent with the flux observed before the periastron passage.

The period of 55 -- 108~days after the periastron was marked by a rise of the GeV flux up to the level $\sim 2\cdot 10^{-6}$~ph/cm$^2$/s on weekly timescales comparable to the values observed during the flare period in 2014 (see top panel of Figure \ref{fig:lc}). Similar to previous periastron passages \cite{Chernyakova20_psrb}, the GeV spectrum during this period was well described by a super-exponential cut-off power-law model (see Tab.~\ref{tab:best_fit_fermi}). After this period the \flat light curve is characterized by a gradual decay of the flux.

Regular observations using SALT during 2021 provided the most frequent observations yet obtained, from 22 days before to 80 days after periastron. Given the marked difference in the X-ray and GeV behavior of the source, the optical observations were remarkably closely aligned with previous observations. A comparison to the behavior in 2014 shows a slightly weaker equivalent width during 2021. Most notably, there does not appear to be any remarkable change in the equivalent width around the period of the increased GeV activity, 50\,days after periastron. This indicates the limits in using the H\,$\alpha$ to directly trace the GeV behavior as suggested by \citep{chernyakova15,Chernyakova2017}. The increased gamma-ray activity has occurred much later during 2021, at a larger binary separation (2.6\,AU at 30 days, 4 AU at 55 days) and the region of the circumstellar disk producing the H\,$\alpha$ emission may be far less effected during this periastron. Future observations at infrared wavelengths may be able to trace the disk's behavior at later orbital phases \citep[see e.g.][and references therein]{klement17}.

Following the model proposed by \citep{Chernyakova20_psrb} we suggest that the X-ray, GeV and TeV emission is generated within the ``emission cone'' formed by the interaction of the pulsar/Be star outflows.

The lower peak X-ray flux during the disk crossings, as well as the slightly lower H\,$\alpha$ equivalent width, may indicate the Be star's circumstellar disk was less dense in 2021, which results in the stand-off shock being further from the pulsar and, consequently, a weaker magnetic field in the emission region. 
The sparser state of the Be star disk during 2021 periastron passage, along with likely a less dense polar outflow, results in a much larger opening angle of the emission cone than what was observed during previous periastron passages.

\begin{figure}
 \centering
 \includegraphics[width=0.35\textwidth]{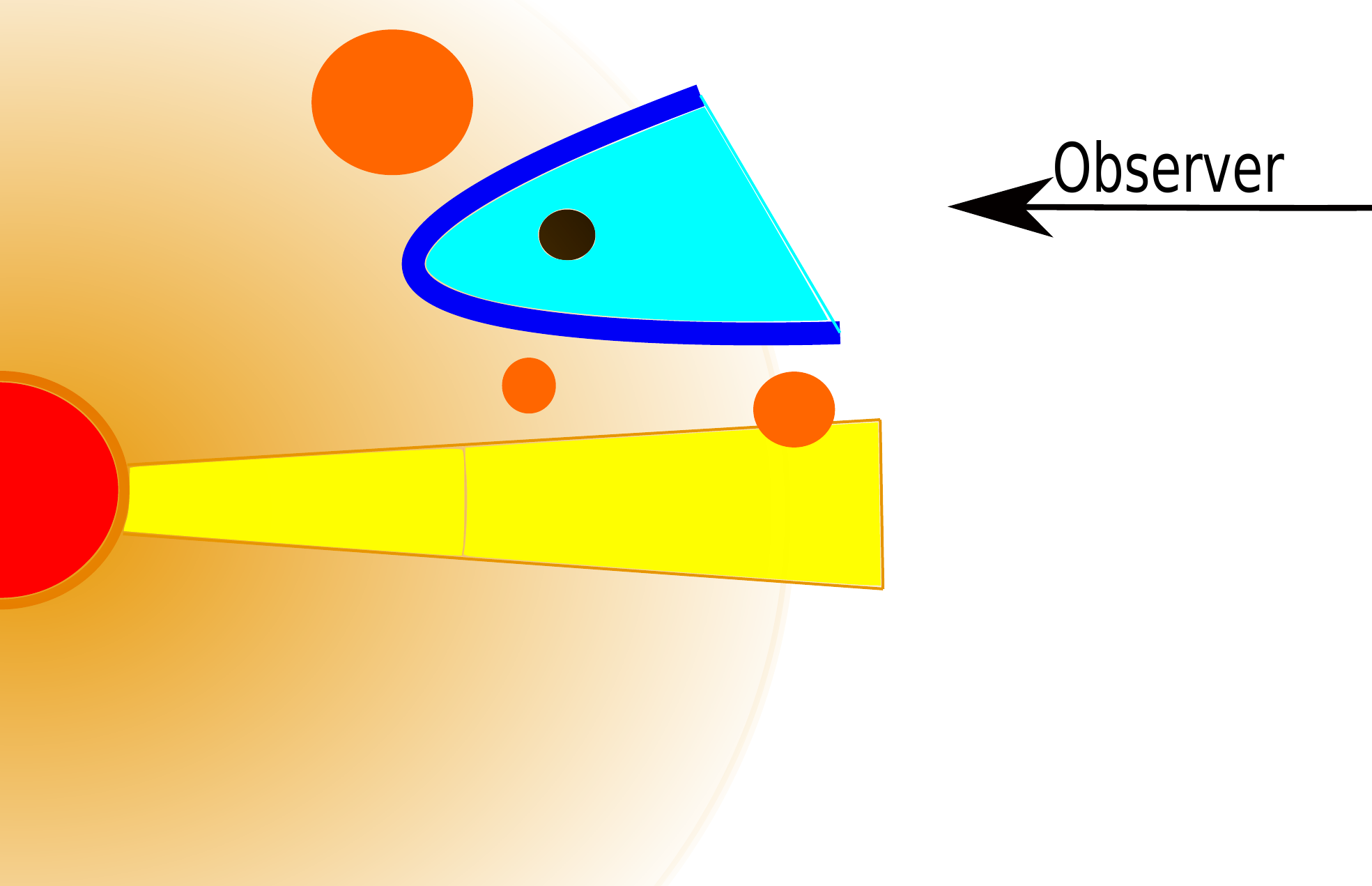}
 \includegraphics[width=0.35\textwidth]{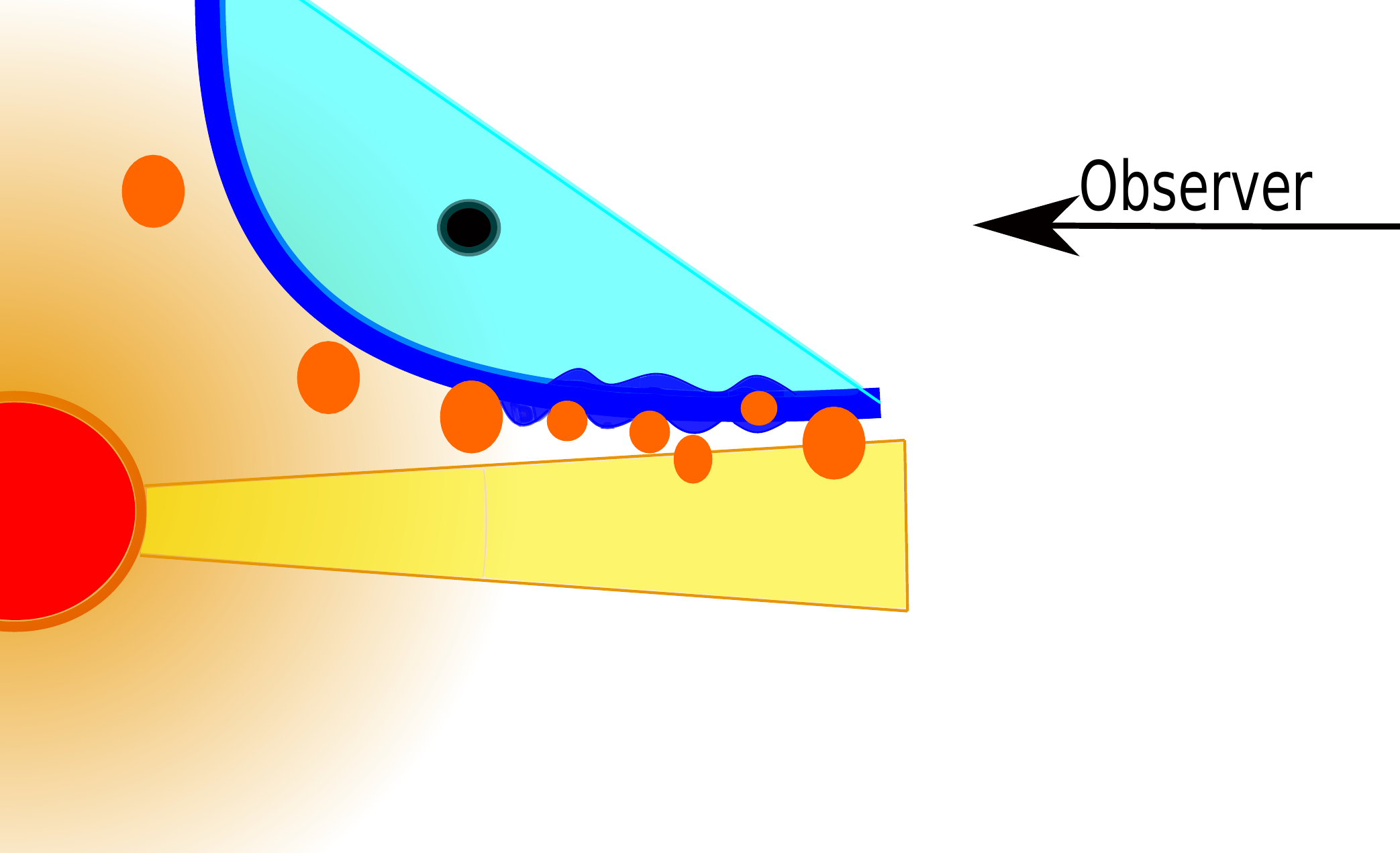}
 \caption{Schematic sketch of the 2017 (left) and 2021 (right) periastron passages. The position of the star and the pulsar are shown with red and black circles, respectively. The equatorial stellar disk is depicted by the yellow shaded regions. The orange shaded regions and the orange circles depict the polar outflow and the clumps in the Be star's disk, respectively. The cyan regions correspond to the unshocked electrons of the pulsar wind. The blue region corresponds to the border of the emission cone along which strongly/weakly shocked electrons of the pulsar wind are leaving the system (see text for more details). All lengths in the figure are not to scale.}
 \label{fig:2017_vs_2021}
\end{figure}

The third peak of the X-ray emission can be interpreted as originating from the presence of a large number of clumps above/below the disk, see right panel of Fig.~\ref{fig:2017_vs_2021}. Such clumps can substantially modify the smooth flow of the strongly shocked relativistic electrons along the emission cone's surface. This effectively increases their escape time from the system and leads to the enhanced level of X-ray emission. The GeV emission in the model of \citep{Chernyakova20_psrb} is connected to the emission from the unshocked electrons of the pulsar wind. These electrons propagate mainly in the inner regions of the emission cone and are thus not significantly affected by the presence of clumps.

The peak level of the GeV emission in the discussed model is proportional to the cone opening angle which naturally explains the relatively low average flux level seen by \flat, see Figure \ref{fig:fermi_lc}(right panel) at shortest time scales.
While at 1~day time scale the highest observed GeV flux at a level of $5\cdot 10^{-6}$~ph/cm$^2$/s is comparable to the level of similar flares seen after the 2017 periastron, we would like to note the substantial difference of the flux variability on intra-day timescales during the 2021 and 2017 passages. In 2017, \psrb demonstrated variability on much shorter (15~mins -- 3~h) timescales, with the flux exceeding the 1~day average by a factor up to $30$ for 15~mins flares \cite{FERMI_PSRB2018}. During the current periastron passage, the source also showed a number of outbursts on a short time scale (5~min -- 3~h), with a flux $\sim 18\cdot 10^{-6}$~erg/cm$^2$/s, see Fig.~\ref{fig:fermi_lc}. Although such outbursts require the luminosity of the source to be at a level of $4.6\cdot 10^{36}$~erg/s which exceeds the spin-down luminosity (assuming $L_{sd}=8.2\cdot 10^{35}$~erg/s) of \psrb by a factor of $\sim 5-6$, this factor is significantly smaller than the factor of $\sim 30$ required to explain the short-term variability of \psrb seen in 2017. We therefore argue that the observed short timescale variability, within the model of~\citep{Chernyakova20_psrb}, is consistent with  a large ($\sim \pi$) opening angle of the emission cone.

The 1-day GeV-band light curve shown in the Fig.~\ref{fig:fermi_lc} additionally indicates flux variability on $0.5-2$ days timescales. 
The observed variability of GeV emission at short (few minutes -- few days) time scales in the model of \citep{Chernyakova20_psrb} can be explained by bremsstrahlung emission from the clumps of the Be star wind entering the emission cone. In this case, the variability timescale corresponds either to the characteristic size of a clump (e.g. shortest time scales) or to the lifetime of smaller clumps in the system (e.g. longest time scales).

In addition to the rapid flares caused by bremsstrahlung, some level of slowly time-varying GeV emission can originate from IC emission of weakly shocked electrons of the pulsar wind. A similar origin was proposed by~\citep{Chernyakova20_psrb} to explain the average GeV flare during the 2017 passage.

\section{Conclusions}

In this paper, we present the first results from intensive multi-wavelength observations of the 2021 periastron passage of \psrb. This periastron was marked by a number of unique features, namely:
\begin{itemize}
 \item A lower X-ray flux during the periods of pre- and post-periastron disk crossings.
 \item The presence of a third X-ray flux peak starting about 30 days after the periastron.
 \item A correlation between the X-ray and radio fluxes during the second X-ray peak, and an absence of such a correlation with the third rise of the X-ray flux.
 \item A substantial delay in the rise of the GeV emission, which started only 55 days after the periastron.
\item A surprising similarity in the variability of the $H_\alpha$ equivalent width compared to previous periastra passages.
\end{itemize}

 We argue that the observed properties can be explained within the model of \cite{Chernyakova20_psrb} under the assumption that the outer parts of the Be star's disk are characterized by lower densities (in comparison to previous periastron passages). We encourage IR observations of this system during the next periastron passages, as the state of the outer parts of the disk is poorly traced by the $H_\alpha$ line variability.
 
A more detailed analysis of the multi-wavelength data from this periastron passage is currently ongoing and will be the subject of an  upcoming publication.

\authorcontributions{Radio data observations and analysis, ChS, ShOS, SMcK, JG, MK; optical data analysis, BvS and IM; X-ray data analysis DM and ST; GeV data analysis, DM; PI of SALT observations, BvS; PI of radio and X-ray proposals, MC; writing, MC, DM, BvS,ShOS; writing---review and editing, AS, GP, SMcK, JG, MK.}


\funding{SMcK was funded by ESA Prodex grant C4000120711. JG acknowledge financial support of CfAR. DM work was supported by DFG through the grant MA 7807/2-1 and DLR through the grant 50OR2104. ST acknowledges financial support by the grant 14.W03.31.0021 of the Ministry of Science and Higher Education of the Russian Federation.}




\dataavailability{X-ray and GeV data are available in a publicly accessible repository. Radio and optical data are available on request.} 

\acknowledgments{This paper uses observations made at the South African Astronomical Observatory (SAAO). This paper uses observation Some of the observations reported in this paper were obtained with the Southern African Large Telescope (SALT) under program 2018-1-MLT-002 (PI: B. van Soelen). 
The Australia Telescope Compact Array is part of the Australia Telescope National Facility which is funded by the Australian Government for operation as a National Facility managed by CSIRO. We acknowledge the Gomeroi people as the traditional owners of the Observatory site.
The authors acknowledge support by the state of Baden-W\"urttemberg through bwHPC.}

\conflictsofinterest{The authors declare no conflict of interest. The funders had no role in the design of the study; in the collection, analyses, or interpretation of data; in the writing of the manuscript, or in the decision to publish the~results.}




\reftitle{References}


\input{journals.tex}
\externalbibliography{yes}
\bibliography{references.bib}


%


\end{paracol}
\end{document}

%% file: journals.tex
\def\aj{AJ}%
\def\actaa{Acta Astron.}%
\def\araa{ARA\&A}%
\def\apj{ApJ}%
\def\apjl{ApJ}%
\def\apjs{ApJS}%
\def\ao{Appl.~Opt.}%
\def\apss{Ap\&SS}%
\def\aap{A\&A}%
\def\aapr{A\&A~Rev.}%
\def\aaps{A\&AS}%
\def\azh{AZh}%
\def\baas{BAAS}%
\def\bac{Bull. astr. Inst. Czechosl.}%
\def\caa{Chinese Astron. Astrophys.}%
\def\cjaa{Chinese J. Astron. Astrophys.}%
\def\icarus{Icarus}%
\def\jcap{J. Cosmology Astropart. Phys.}%
\def\jrasc{JRASC}%
\def\mnras{MNRAS}%
\def\memras{MmRAS}%
\def\na{New A}%
\def\nar{New A Rev.}%
\def\pasa{PASA}%
\def\pra{Phys.~Rev.~A}%
\def\prb{Phys.~Rev.~B}%
\def\prc{Phys.~Rev.~C}%
\def\prd{Phys.~Rev.~D}%
\def\pre{Phys.~Rev.~E}%
\def\prl{Phys.~Rev.~Lett.}%
\def\pasp{PASP}%
\def\pasj{PASJ}%
\def\qjras{QJRAS}%
\def\rmxaa{Rev. Mexicana Astron. Astrofis.}%
\def\skytel{S\&T}%
\def\solphys{Sol.~Phys.}%
\def\sovast{Soviet~Ast.}%
\def\ssr{Space~Sci.~Rev.}%
\def\zap{ZAp}%
\def\nat{Nature}%
\def\iaucirc{IAU~Circ.}%
\def\aplett{Astrophys.~Lett.}%
\def\apspr{Astrophys.~Space~Phys.~Res.}%
\def\bain{Bull.~Astron.~Inst.~Netherlands}%
\def\fcp{Fund.~Cosmic~Phys.}%
\def\gca{Geochim.~Cosmochim.~Acta}%
\def\grl{Geophys.~Res.~Lett.}%
\def\jcp{J.~Chem.~Phys.}%
\def\jgr{J.~Geophys.~Res.}%
\def\jqsrt{J.~Quant.~Spec.~Radiat.~Transf.}%
\def\memsai{Mem.~Soc.~Astron.~Italiana}%
\def\nphysa{Nucl.~Phys.~A}%
\def\physrep{Phys.~Rep.}%
\def\physscr{Phys.~Scr}%
\def\planss{Planet.~Space~Sci.}%
\def\procspie{Proc.~SPIE}%
\let\astap=\aap
\let\apjlett=\apjl
\let\apjsupp=\apjs
\let\applopt=\ao